\documentclass[12pt]{article}

\usepackage{amssymb}
\usepackage[dvips]{graphicx}

\newcommand {\beq}{\begin{equation}}
\newcommand {\eeq}{\end{equation}}
\newcommand {\beqa}{\begin{eqnarray}}
\newcommand {\eeqa}{\end{eqnarray}}
\newcommand {\n}{\nonumber \\}

\def\pa{\partial}

\begin{document}
\setlength{\oddsidemargin}{0cm}
\setlength{\baselineskip}{7mm}

\begin{titlepage}
\renewcommand{\thefootnote}{\fnsymbol{footnote}}
\begin{normalsize}
\begin{flushright}
\begin{tabular}{l}
October 2012
\end{tabular}
\end{flushright}
  \end{normalsize}

~~\\

\vspace*{0cm}
    \begin{Large}
       \begin{center}
         {Comment on ``ADM reduction of IIB on ${\cal H}^{p,q}$ to dS braneworld"}
       \end{center}
    \end{Large}
\vspace{1cm}

\begin{center}
           Matsuo S{\sc ato}$^{1)}$\footnote
            {
e-mail address : 
msato@cc.hirosaki-u.ac.jp}
           {\sc and}
           Asato T{\sc suchiya}$^{2)}$\footnote
           {
e-mail address : satsuch@ipc.shizuoka.ac.jp}\\
      \vspace{1cm}
       
        $^{1)}$ {\it Department of Natural Science, Faculty of Education, Hirosaki University}\\
               {\it Bunkyocho 1, Hirosaki, Aomori 036-8560, Japan}\\
      \vspace{0.5cm}
                    
        $^{2)}$ {\it Department of Physics, Shizuoka University}\\
                {\it 836 Ohya, Suruga-ku, Shizuoka 422-8529, Japan}
\end{center}

\hspace{5cm}

\begin{abstract}
\noindent
We make a comment on ``ADM reduction of IIB on ${\cal H}^{p,q}$ to dS braneworld" by
E. Hatefi, A. J. Nurmagambetov and I. Y. Park, arXiv:1210.3825.
\end{abstract}
\vfill
\end{titlepage}
\vfil\eject

\setcounter{footnote}{0}

\renewcommand{\thefootnote}{\arabic{footnote}} 
In \cite{Sato:2002kv,Sato:2003ky,Sato:2003uc},
we showed that the effective action of D-brane, the Born-Infeld action with
the Wess-Zumino terms, is a solutions to
the Hamilton-Jacobi equation in supergravity. 
In \cite{Sato:2004ic} we also generalized this to the case of M2-brane and M5-brane.
To derive the Hamilton-Jacobi equation where the radial direction is regarded as the time direction,
we made a consistent truncation of supergravity over $S^{8-p}$ in the case of the $p$-brane
and obtained a $p+2$-dimensional theory.
Here, as usual, the consistent truncation means that all the classical solutions of the truncated theory are
also classical solutions of the original theory.

Recently, the authors of \cite{Hatefi:2012bp} have studied a similar Hamilton-Jacobi equation 
in a five-dimensional theory.
In obtaining the five-dimensional theory from ten-dimensional supergravity by a consistent
truncation over ${\cal H}^5$ or $S^5$ and 
deriving the Hamilton-Jacobi equation in the five-dimensional theory,
they use our calculation for the $p=3$ case in \cite{Sato:2002kv} as a reference. However, they claim that
there is an error in our consistent truncation over $S^5$. They obtained a five-dimensional theory
and a Hamilton-Jacobi equation that differ from ours.
They show that the effective action of D3-brane is not a solution
to their Hamilton-Jacobi equation.

In this brief article, by deriving our truncated theory from a different viewpoint,
we verify that our results in \cite{Sato:2002kv} are correct.
We also point out an error in \cite{Hatefi:2012bp}.
More specifically, the truncated theory 
obtained in the Einstein frame in \cite{Hatefi:2012bp} is consistent with ours,
but there is an error in transforming this truncated theory
to the one in the string frame.

As in \cite{Sato:2002kv,Hatefi:2012bp}, we consider ten-dimensional type IIB supergravity.
The discrepancy between \cite{Sato:2002kv}
and \cite{Hatefi:2012bp} exists for the part of the metric and the dilaton. 
Also, ignoring all the fields other than the metric and the dilaton is consistent.
We therefore concentrate on the metric and the dilaton throughout this article. 
Then, the ten-dimensional action in the string frame takes the form
\beqa
I_{10}=\frac{1}{2\kappa_{10}^{\:2}} \int d^{10} X \sqrt{-G}
e^{-2\Phi} \left(R_G+4\pa_M \Phi \pa^M \Phi \right).
\label{10Daction}
\eeqa
In order to make a consistent truncation over $S^5$, 
we split the ten-dimensional
coordinates $X^M$ into two parts, as $X^M=(\xi^{\alpha},\; \theta_i)
\;\;(\alpha=0,\cdots,4, \;\; i=1,\cdots,5)$, where the $\xi^{\alpha}$ are
five-dimensional coordinates and the $\theta_i$ parametrize
$S^5$. We make the following ansatz for 
the metric and the dilaton: 
\beqa
ds_{10}^{\:2}&=&G_{MN}dX^MdX^N \n
&=& h_{\alpha\beta}(\xi) \: d\xi^{\alpha}d\xi^{\beta}
+e^{\rho(\xi)/2} \: d\Omega_5, \nonumber\\
\Phi &=& \phi (\xi),
\label{ansatz}
\eeqa
where $h_{\alpha\beta}$ is a five-dimensional metric, and
$d\Omega_5$ is the $SO(6)$-invariant metric of unit $S^5$.
We set the other modes to zero.

(\ref{ansatz}) is the most general form of the metric and the dilaton
that is invariant under the $SO(6)$ transformation. Namely, we discard all the $SO(6)$-variant modes.
(\ref{10Daction}) is $SO(6)$-invariant so that the terms in (\ref{10Daction}) including these $SO(6)$-variant modes
are at least quadratic order in those modes. Hence, 
the equations of motion for these $SO(6)$-variant modes
are automatically satisfied under (\ref{ansatz}). 
This implies that the desired five-dimensional
theory can be obtained by substituting (\ref{ansatz}) into (\ref{10Daction}) 
(see, for instance, argument in \cite{Kawai:2010sf})\footnote{This procedure is still valid
when the Kalb-Ramond field
and the Ramond-Ramond fields except the self-dual 4-form are included and
the $SO(6)$ invariant ansatz is made for those fields.}.
Thus, by substituting (\ref{ansatz}) into (\ref{10Daction}) and
using the formula in appendix, we obtain the five-dimensional theory
\begin{eqnarray}
I_5 = \frac{1}{2\kappa_5^2} \int d^5 \xi \sqrt{-h} \left[
e^{-2\phi+\frac{5}{4}\rho}
\left( R^{(5)}+4\partial_{\alpha}\phi\partial^{\alpha}\phi 
+\frac{5}{4}\partial_{\alpha}\rho\partial^{\alpha}\rho
-5\partial_{\alpha}\phi\partial^{\alpha}\rho\right) + e^{-2\phi+\frac{3}{4}\rho}R^{(S^5)}\right] ,
\label{5Daction}
\end{eqnarray}
where 
\beqa
\frac{1}{2\kappa_5^2}=\frac{\mbox{volume of }S^5}{2\kappa_{10}^{\:2}}, \;\;\; 
R^{(S^5)}=20.
\nonumber
\eeqa
This indeed agrees with the five-dimensional theory in \cite{Sato:2002kv} with
the Kalb-Ramond field and the Ramond-Ramond fields set to zero.
Thus, our consistent truncation in \cite{Sato:2002kv} is correct\footnote{We have verified that 
the equations of motion of (\ref{5Daction}) reproduce 
the equations obtained by substituting (\ref{ansatz}) into the equations of motion of (\ref{10Daction}). This is another check for
the correctness of (\ref{5Daction}). The claim in footnote 4 of \cite{Hatefi:2012bp} that this is not the case is wrong.}.

Let us see that the above result is consistent with (2.6) of \cite{Hatefi:2012bp}.
The authors of \cite{Hatefi:2012bp} work in the Einstein frame where ten-dimensional action
takes the form
\beqa
I_{10}=\frac{1}{2\kappa_{10}^{\:2}} \int d^{10} X \sqrt{-G}
\left(R_G-\frac{1}{2}\pa_M \Phi \pa^M \Phi \right).
\label{10DactionEinstein}
\eeqa
This action is obtained from (\ref{10Daction}) by making a replacement
\beqa
G_{MN}\rightarrow e^{\frac{1}{2}\Phi}G_{MN}.
\label{Einstein frame}
\eeqa
They make the following ansatz for this new metric and the dilaton:
\beqa
ds_{10}^{\:2}
&=& e^{2\tilde{\rho}(\xi)}\tilde{h}_{\alpha\beta}(\xi) \: d\xi^{\alpha}d\xi^{\beta}
+e^{-6\tilde{\rho}(\xi)/5} \: d\Omega_5, \nonumber\\
\Phi &=& \phi (\xi).
\label{ansatz2}
\eeqa
$\mbox{}$From (\ref{ansatz}), (\ref{Einstein frame}) and (\ref{ansatz2}), we find the relationship 
\beqa
h_{\alpha\beta}&=&e^{\frac{1}{2}\phi+2\tilde{\rho}}\tilde{h}_{\alpha\beta}, \n
\rho&=& -\frac{12}{5}\tilde{\rho}+\phi.
\label{relationship}
\eeqa
By substituting (\ref{relationship}) into (\ref{5Daction}) and using the formula in appendix, we obtain
\begin{eqnarray}
I_5 = \frac{1}{2\kappa_5^2} \int d^5 \xi \sqrt{-\tilde{h}} \left[
\tilde{R}^{(5)}-\frac{1}{2}\partial_{\alpha}\phi\partial^{\alpha}\phi 
-\frac{24}{5}\partial_{\alpha}\tilde{\rho}\partial^{\alpha}\tilde{\rho}
+ e^{\frac{16}{5}\tilde{\rho}}R^{(S^5)}\right] .
\label{5Daction2}
\end{eqnarray}
This agree with (2.6) of \cite{Hatefi:2012bp} with
the Kalb-Ramond field and the Ramond-Ramond fields set to zero.

In deriving the Hamilton-Jacobi equation,
the authors of \cite{Hatefi:2012bp} move to the string frame\footnote{The definition of the string frame
in \cite{Hatefi:2012bp} is
different from ours.} by making a replacement:
\beqa
\tilde{\rho}&=&-\frac{5}{12}\check{\rho}, \n
\tilde{h}_{\alpha\beta}&=&e^{-\frac{1}{2}\phi+\frac{5}{6}\check{\rho}}\check{h}_{\alpha\beta}.
\eeqa
Then, (\ref{5Daction2}) leads to
\beqa
I_5 &=& \frac{1}{2\kappa_5^2} \int d^5 \xi \sqrt{-\check{h}} \left[e^{-\frac{3}{4}\phi+\frac{5}{4}\check{\rho}}
\left(\check{R}^{(5)}-\frac{10}{3}\check{\nabla}^2\check{\rho}
-\frac{35}{12}\partial_{\alpha}\check{\rho}\partial^{\alpha}\check{\rho}
+2\check{\nabla}^2\phi-\frac{5}{4}\partial_{\alpha}\phi\partial^{\alpha}\phi
+\frac{5}{2}\partial_{\alpha}\phi\partial^{\alpha}\check{\rho}\right) \right.\n
&&\left.\qquad\qquad\qquad\qquad+ e^{-\frac{5}{4}\phi+\frac{3}{4}\check{\rho}}R^{(S^5)}\right].
\eeqa
However, the term $\frac{5}{2}\partial_{\alpha}\phi\partial^{\alpha}\check{\rho}$ is missing in
(3.2) of \cite{Hatefi:2012bp}. Thus, (3.2) of \cite{Hatefi:2012bp} is unfortunately incorrect.
The Hamilton-Jacobi equation derived from (3.2) of \cite{Hatefi:2012bp} is of course different from ours.
Hence, their claim that the D3-brane effective action is not a solution to the
Hamilton-Jacobi equation is wrong.


\section*{Acknowledgements}
We would like to thank T. Shiromizu for informing us of \cite{Hatefi:2012bp}.
The work of A.T. is supported in part by Grant-in-Aid
for Scientific Research
(No. 24540264, and 23244057)
from JSPS.


\section*{Appendix: Some useful formulae}
First, we write down the Wely transformation in $D$ dimensions.
Under $\tilde{g}_{\mu\nu}=e^{2\omega}g_{\mu\nu}$,
\beqa
\tilde{R}=e^{-2\omega}(R-2(D-1)\nabla^2\omega-(D-1)(D-2)\partial_{\mu}\omega \partial^{\mu}\omega).
\eeqa
Next, we consider the following reduction of the ten-dimensional space-time 
on $S^{8-p}$:
\beqa
ds_{10}^{\:2}&=&G_{MN} \: dX^M dX^N \n
&=& h_{\alpha\beta}(\xi) \: d\xi^{\alpha}d\xi^{\beta} 
+e^{\sigma(\xi)/2} \: d\Omega_{8-p}.
\eeqa
Here, the $\xi^{\alpha}$ are $(p+2)$-dimensional coordinates, and $S^{8-p}$ is
parametrized by $\theta_1,\cdots,\theta_{8-p}$. The ten-dimensional curvature
is represented by the $(p+2)$-dimensional curvature
and the $(8-p)$-dimensional
curvature as
\beqa
R_G=R^{(p+2)}-\frac{8-p}{2}\nabla^{(p+2)}_{\alpha}\nabla^{(p+2)\alpha}\sigma
-\frac{(8-p)(9-p)}{16}\pa_{\alpha}\sigma \: \pa^{\alpha}\sigma
+e^{-\sigma/2} \: R^{(S^{8-p})},
\eeqa
where $R^{(S^{8-p})}$ is the constant curvature of $S^{8-p}$.

\end{document}